\documentclass[conference]{IEEEtran}
\usepackage{authblk,amsmath,amsfonts,amssymb,epsfig,color}

\usepackage{cite,footnote,authblk}

\newcommand{\pr}[1]{\left( #1\right)}

\newcommand{\es}[1]{\begin{equation}\begin{split}#1\end{split}\end{equation}}

\newcommand{\R}{\mathbb{R}}

\newcommand{\dd}{\textrm{d}}

\begin{document}
\title{Maximum Likelihood based Multihop Localization in Wireless Sensor Networks}
\author[1]{CamLy Nguyen}
\author[2]{Orestis Georgiou}
\author[1]{Yusuke Doi}
\affil[1]{Network System Laboratory, Corporate Research \& Development Center, Toshiba Corporation,}
\affil[ ]{1 Komukai-Toshiba-cho, Saiwai-ku, Kawasaki 212-8582, Japan}
\affil[2]{Toshiba Telecommunications Research Laboratory, 32 Queens Square, Bristol, BS1 4ND, UK}
\maketitle

\begin{abstract}
For data sets retrieved from wireless sensors to be insightful, it is often of paramount importance that the data be accurate and also location stamped.
This paper describes a maximum-likelihood based multihop localization algorithm called kHopLoc for use in wireless sensor networks that is strong in both isotropic and anisotropic network deployment regions.
During an initial training phase, a Monte Carlo simulation is utilized to produce multihop connection density functions. 
Then, sensor node locations are estimated by maximizing local likelihood functions of the hop counts to anchor nodes. 
Compared to other multihop localization algorithms, the proposed kHopLoc algorithm achieves higher accuracy in varying network configurations and connection link-models.
\end{abstract}


\section{Introduction \label{sec:intro}}


Wireless sensor networks (WSN) are composed of a set of spatially distributed wireless nodes, with sensing and transceiving capabilities, tasked with monitoring physical or environmental conditions, such as temperature, sound, pressure, radiation, etc. 
Data collected is then wirelessly passed through the network to a main gateway for storing and processing.
WSNs are essential for applications such as environmental monitoring, target tracking, disaster relief and rescue operations \cite{tilak2002taxonomy,akyildiz2010wireless}. 
Nowadays, they are also becoming an indispensable part of smart technologies with applications in smart cities and smart buildings \cite{kortuem2010smart}.

In view of the internet of things future vision, almost every device will soon have transceiving capabilities, be packed with sensors, connected to a network and producing huge data sets.
As is often the case however, location information is vital for the insightful processing of this data.
GPS modules may be embedded to each sensor node enabling it to autonomously discover its location both accurately and on demand, however this does not come without a cost to the manufacturer (and hence the user) and the node's power source - about 30mA at 3.3V. 
Moreover, in some extreme instances such as sand storms and blizzards, or simply when operating indoors, satellite signals cannot reach the sensor nodes.

To alleviate such problems, cooperative schemes have been developed to estimate the locations of sensor nodes with the assistance of nodes which have perfect location information \cite{patwari2005locating,wymeersch2009cooperative}. 
The nodes whose locations are known and the nodes whose locations are unknown are usually called \emph{anchor nodes} and \emph{target nodes} respectively and the localization techniques can be broadly classified into two schemes: \textit{range-based} and \textit{range-free} schemes. 
Range-based schemes \cite{dil2006range} assume that the distance or angle between anchor nodes and target node can be measured based on signal measurements such as received signal strength indication (RSSI), time of arrival (TOA), or angle of arrival (AOA). 
In large-scale WSNs where signal range is limited however, range based schemes typically require a lot of anchor nodes to produce accurate results.
On the other hand, range-free schemes \cite{he2003range} estimate inter-node distances based on hop count information, thus all target nodes can be localized with fewer anchor nodes.

Conventional range-free approaches \cite{niculescu2003dv,wu2011improved,chen2012improved} usually consider isotropic WSNs where sensor nodes are uniformly distributed in a regular region (e.g. a square domain), thus the distance between two nodes is assumed to be proportional to their hop count.
The celebrated DV-hop algorithm \cite{niculescu2003dv} estimates the average one-hop distance, and then multiplies this by the hop count to at least three anchor nodes before trilaterating. 
Although improved DV-hop algorithms have been suggested  in the literature \cite{wu2011improved,chen2012improved}, performance gains have been limited, even more so when used in anisotropic WSNs where factors such as irregular radio propagation, obstacles, nonuniform node distributions degrade the hop-distance proportionality assumption. 
DV-hop-like variants which are anisotropic network compatible have also been proposed \cite{
xiao2010reliable,liu2011anchor}, the main idea usually being to reduce the estimation error by reducing the effect of unreliable anchors. 
Thus, these algorithms require an increased number of anchor nodes, possibly as many as range-based schemes. 
Other approaches \cite{lee2014multihop, wang2009range} attempt to use approximate shortest paths to reduce the effect of anisotropic networks, yet \cite{wang2009range} under performs in irregular-shaped regions as shown in \cite{lee2014multihop} which comes with large communication and computational overhead.

In this paper, we propose a \emph{maximum likelihood based multihop localization algorithm} called \emph{kHopLoc} which achieves good performance in both isotropic and anisotropic WSNs. 
The algorithm first runs a training phase during which a Monte Carlo simulation is utilized to produce accurate multihop connection probability density functions (described later).
In its second phase, the algorithm constructs likelihood functions for each target node based on their hop counts to all reachable anchor nodes which it then maximizes to produce localization information. 
Unlike most DV-hop algorithms which use only first order statistics, the proposed kHopLoc algorithm generates and uses the full multihop density distributions (even for anisotropic networks) thus constructing accurate likelihood functions and in turn localization results. 
In addition, our algorithm's communication cost is about half of most DV-hop-like algorithms and computational cost is much smaller.

The rest of this paper is organized as follows. In Section \ref{network-model} we describe the network and system model. 
In Sec. \ref{algorithm} we present the details of our proposed kHopLoc algorithm. 
In Sec. \ref{performance} we evaluate the performance of kHopLoc through numerical simulations. 
Finally, Sec. \ref{conc} concludes the paper with a summary and some discussion on future work plans.

\section{Network Definitions and System Model 
\label{network-model}}

Consider a WSN of $N$ total sensor nodes, $M$ of which are anchor nodes (i.e. have perfect location information). 
The remaining $N-M$ sensors are target nodes (i.e. locations are unknown).
All nodes are randomly distributed in some subset of $\R^2$ and are equipped with isotropic antennas. 
Due to phenomena such as fading and multi path, the communication model adopted may not be well described by the simplistic disk model (where two nodes connect if they are within a finite range of each other).
Instead, we adopt a \textit{random connection model} where nodes connect with a distance dependent probability while also accommodating for environmental parameters such as path loss exponent \cite{coon2012full}. 
Namely, we consider two such communication models in order to show that the kHopLoc algorithm works in different communication models: \textit{Rayleigh fading} communication model and \textit{Quasi Unit Disk Graph} (QUDG) communication model \cite{gao2011hop}. 
The latter model also allows the direct comparison of kHopLoc with that given in \cite{lee2014multihop}.
Moreover, we consider isotropic networks in which the sensor nodes are distributed in a square, and anisotropic networks where sensor nodes are deployed in irregular shaped regions. 

\subsection{Rayleigh fading communication model}

The \emph{pair connectedness function} $H$ defined as the probability that two nodes are directly connected. 
One way of formulating $H$ is thus the complement of the information outage probability with respect to a mutual information threshold $\vartheta$. 
For a narrow band transmission subject to small-scale fading
\es{
H &= P\pr{ \log_2 (1+\mathrm{SNR} \cdot X) >   \vartheta }
= F_X \pr{ \frac{2^{\vartheta}-1}{\mathrm{SNR}} } \label{connectednessfunction}
,}
where $X$ is a random variable defining the gain of the wireless channel between the two nodes, $F_X$ is its complementary cumulative distribution function (CCDF), and SNR is the long-term average received signal-to-noise ratio. Since the nodes are equipped with isotropic antennas, the $\mathrm{SNR} \propto d^{-\eta}  \label{SNR}$, where $d$ is the distance between the two nodes, and $\eta$ is the path loss exponent. 
Adopting a Rayleigh fading model for the small-scale fading gain $X\sim\exp(1)$ implies that the CCDF of $X$ is written as 
$F_X(x)= \exp(-x)$, and the pair connectedness function can be expressed as
\es{
H= \exp(-\beta d^\eta)
\label{connectedness}
,}
where $\beta$ is a constant depending on the node transmission power, wavelength, information threshold etc. and can be understood as an effective communication range $r_0=\beta^{-1/\eta}$.
Notice that in the theoretical limit of $\eta\to\infty$, we have that $r_0\to 1$ and $H$ converges to the unit disk model.
Without loss of generality, we will henceforth set $\beta=1$.

\subsection{Quasi Unit Disk Graph communication model}

In the QUDG communication model \cite{gao2011hop}, the probability $H$ that two nodes a distance $d$ apart are directly connected is 
\es{
H= \begin{cases}
1  & \text{if } d < d_{max}/DOI \\
\frac{DOI(d_{max}-d)}{d_{max}(DOI-1)}       & \text{if } d \in [d_{max}/DOI, d_{max} ] \\
0  & \text{if } d > d_{max}/DOI
  \end{cases}
\label{QUDG}}
where $d_{max}>0$ and $DOI > 1$ are the maximum successful transmission distance and the degree of radio irregularity respectively.   

\begin{figure}[t]
\centering
\includegraphics[width= 8.5cm]{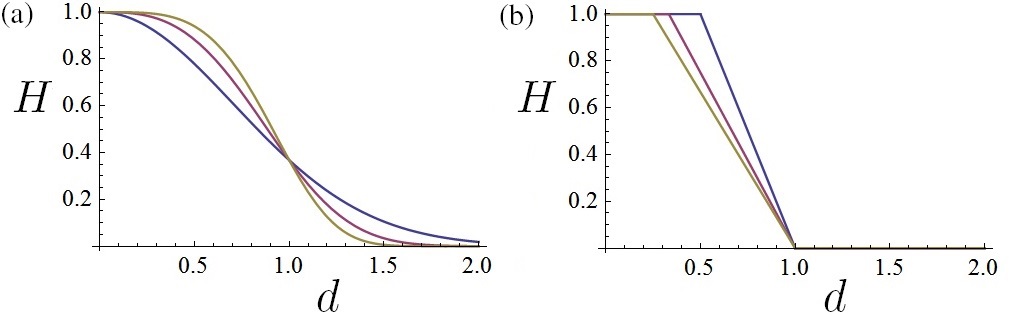}
\caption{Plots of the pair connectedness function $H$ using (a) the Rayleigh fading communication model \eqref{connectedness} for parameters $\beta=1$ and $\eta=2,3,4$, and (b) the Quasi Unit Disk Graph communication model \eqref{QUDG} for parameters $d_{max}=1$ and $DOI=2,3,4$.}
\label{fig:hfunctions}
\end{figure}

During computer simulations, the network graph edges are formed if a random number $\zeta\in[0,1]$ is less than the calculated $H$ of the respective communication model.
Two nodes sharing a successful link are called one-hop neighbours. 
More generally, two nodes with the minimum hop count of $k\geq1$ (measured along the shortest path) are called $k$-hop neighbours.
The two $H$ functions are plotted in Fig. \ref{fig:hfunctions}.


\section{kHopLoc algorithm description
\label{algorithm}}


Our proposed kHopLoc algorithm is composed of three simple steps.
In the first step, each target node counts the minimum number of hops to each anchor node.
In the second step, the conditional probability density function is generated for the probability density function $p(d|k)$ of the Euclidean inter-node distance $d$ given a hop count from target to anchor node equal to $k=1,2,\ldots$.
This step is preferably performed centrally using a Monte Carlo simulation after which fitting parameters are flooded through the network.
Finally, in the third step, each target node calculates its own position by maximising the joint conditional density function to all reachable anchors.
We now describe each of these steps in more detail.

\subsection{Step 1: Minimum hop-count to anchor nodes
\label{DV}}

In this initial step, a classic distance vector exchange routine takes place (as in most DV-hop type localization algorithms \cite{niculescu2003dv}) where all target nodes in the network get distances, in hops, to their reachable anchors i.e. at most $K$-hops away. 
This can be initiated by anchors nodes which broadcast beacons to be 
flooded throughout the network containing the anchor's location with a hop-count value initialized to one.
Each node maintains a table containing the coordinates $(x_i, y_i)$ of anchor node $i$ and the minimum hop distance to it $h_i$, and exchanges such updates only with its immediate neighbours.


\subsection{Step 2: Multihop connection probability density function
\label{k-hop-probability}}

The conditional probability density function $p(d|k)$ of inter-node distance $d$ given the minimum hop count $k$, is to the best of our knowledge the necessary ingredient which differentiates our algorithm from all others.
From Bayes theorem  we have that $p(d|k)= p(k|d)p(d)/p(k)$.
We may therefore calculate $p(d|k)$ indirectly through $p(k|d)p(d)$ and $p(k)$. 
Since $p(k)$ is independent of the inter-node distance, it does not affect the maximization calculation and so will be ignored later on. 

The conditional probability of the target-to-anchor hop count being $k$, given that they are a distance $d$ apart is given by $p(k|d)$ and can be approximately formulated in closed form as given by \cite{mao2010probability,directional}.
Similarly, the probability distribution of the distance between two random points $p(d)$ can be formulated as in \cite{Durrani}. 
However, the probability density function in \cite{mao2010probability} consists of multiple integrals and assumes very dense networks, making it difficult to calculate, even more so by the sensor's weak processing unit. 
Therefore, in order to make progress we propose here a method for producing the probability density function $p(k|d)p(d)$ and then flooding the network with the required fitting parameters.
First, Monte Carlo simulations are employed to calculate the discrete values of $p(k|d)p(d)$, which are then fitted into a continuous functions.

\subsubsection{Monte Carlo simulation}
Let $K< N$ and $D\gg 1$ be the maximum allowed hop count and maximum allowed Euclidean distance between a target and anchor node respectively. Roughly, $D$ should be at least larger than $K d_{max}$ or $K/\beta$.
Partition the disk of radius $D$ into $L$ concentric shells (like an archery target) of widths $2\delta=D/L$ indexed by $l=1,2\ldots L$ such that each shell has central radius of $d_l = \delta(2 l +1)$.
We want to calculate the value of $p(k|d_l)p(d_l)$ for each $k=1,2,\ldots K$ and each $d_l$ with $l =1,2\ldots L$.
The Monte Carlo simulation can now be performed at one or more sensor nodes (preferably ones with significant processing power) or a central server e.g. the gateway as follows:
\begin{enumerate}[]
\item Generate random coordinates with intensity $\rho=N/A$ inside some predefined region. 
If the total number of nodes $N$ and the WSN deployment region shape and area $A$ are known then this is easy. 
If the WSN region shape is unknown, then a large square region can be used - it is shown that this does not affect the results significantly (see Fig. \ref{fig:error-Cshape}). 
If the density of nodes is unknown, then it can be estimated as described below in subsection \ref{density-estimation}. 

\item Generate communication links between nodes based on their mutual distances and appropriate connection probability function $H$ (e.g. \eqref{connectedness} or \eqref{QUDG} or other).

\item For each $k\in[1,K]$ and $l\in[1,L]$, calculate the discretized cumulative probability $p(k|d_l)p(d_l)\approx P(k|(d_l-\delta<d \leq d_l+\delta))P(d_l-\delta<d \leq d_l+\delta)/2\delta$. 

\item Repeat steps (1) - (3) above several times in a Monte Carlo fashion in order to refine the estimated $p(k|d_l)p(d_l)$.
\end{enumerate}

\subsubsection{Fitting}
For each $k \in[1,K]$, fit the discrete probability distribution $p(k|d)p(d)$ to the following function of $d$
\es{
p(k|d)p(d)= \exp( -A(k)(d-B(k))^2+ C(k)) 
\label{fitting}
,}
where $A(k), B(k), C(k)$ are functions of $k$, e.g. polynomials of degree $\wp \gg 1$.
In the simulations that follow we use $\wp=4$.
The Gaussianity of \eqref{fitting} was inspired by the extensive simulations results and analysis presented in \cite{zhang2012hop}.

\subsubsection{Density estimation\label{density-estimation}}

This section describes a simple method of estimating node density $\rho$, a prerequisite for performing the Monte Carlo simulations and building the said distributions.
In a uniformly distributed network, the node density is defined as $\rho=N/A$, where $A$ is the WSN deployment region area, thus it can be approximated by the average number of one-hop neighbours $N_e$ divided by it's average communication area $A_e$. 
The former can be simply determined by each sensor counting the number of $1$-hop neighbours from each node while the latter can be calculated 
\es{
A_e= \int_0^{2\pi}\int_0^{\infty}r H(r)\dd r \dd\theta 
\label{area_cal}
,}
where $H(r)$ is the pair connectedness function of two nodes whose relative distance is $r$ (see \eqref{connectedness} and \eqref{QUDG}).
The estimated density $\rho_e$ is then sent to the nearest processor which performs an average to obtain a refined estimate.


\subsection{Step 3: Maximum likelihood based Multihop localization
\label{ML-localization}}

This subsection describes our proposed kHopLoc algorithm utilizing maximum likelihood methods. 
First we introduce the likelihood function and then we address a method to maximize it. 
Consider a target node $X$ and let $h_i$ be the hop count measure along the shortest path from $X$ to anchor node $i \in[1,M]$. 
The likelihood function of $X$ at the yet undetermined coordinate $(x,y)$ is defined as 
\es{
L(x,y)= p(x,y| h_1, h_2, ...h_M)
\label{likelihood1}
.}
It follows that the best estimate of the true location of node $X$ is the value that maximizes it's  likelihood
\es{
(x^*,y^*)= \operatorname*{arg\,max}_{(x,y)} L(x,y)
\label{maxlikelihood}
.}
Assuming that the probability density functions $p(x,y|h_i)$ and $p(x,y|h_j)$ are mutually independent for $i\not=j$, equation \eqref{likelihood1} can be written as
\es{
L(x,y)= \prod_{i=1}^{M} p(x,y| h_i) = \prod_{i=1}^{M} 2 \pi p(d_i| h_i)
\label{likelihood2}
,}
where $d_i= \sqrt{(x-x_i)^2+ (y-y_i)^2}$ and $(x_i, y_i)$ are the coordinates of anchor node $i$.
Invoking Bayes theorem $p(d_i|h_i) = p(h_i|d_i)p(d_i) / p(h_i)$ yields
\es{
 L(x,y) &= (2\pi)^M \prod_{i=1}^{M}  \frac{p(h_i|d_i)p(d_i)}{p(h_i)}
\label{likelihood3}
.}
Substituting equation \eqref{likelihood3} back into equation \eqref{maxlikelihood} , we obtain
\es{
(x^*,y^*)= \operatorname*{arg\,max}_{(x,y)} \frac{\prod_{i=1}^{M}p(h_i|d_i)\cdot \prod_{i=1}^{M}p(d_i)}{\prod_{i=1}^{M}p(h_i)}
\label{maxlikelihood2}
.}
Because $p(h_i)$ is independent of $(x,y)$, the product in the denominator can be eliminated from equation \eqref{maxlikelihood2} such that
\es{
(x^*,y^*) = \operatorname*{arg\,max}_{(x,y)} \prod_{i=1}^{M}p(h_i|d_i)p(d_i)
\label{maxlikelihood3}
.}
Finally, substituting our fitted equation \eqref{fitting} into equation \eqref{maxlikelihood3} gives the result
\es{
(x^*,y^*) &= \operatorname*{arg\,max}_{(x,y)} \prod_{i=1}^{M}\exp(-A(h_i)(d_i-B(h_i ))^2+C(h_i) ) \\
&= \operatorname*{arg\,min}_{(x,y)} \sum_{i=1}^{M} A(h_i)(d_i-B(h_i ))^2
\label{maxlikelihood4}
.}
The right hand side of equation \eqref{maxlikelihood4} can be easily calculated using gradient descent method or Newton method for example and can be performed by each target node independently.

\begin{figure}[t]
\centering
\begin{tabular}{ccc}
\includegraphics[width= 3.5cm]{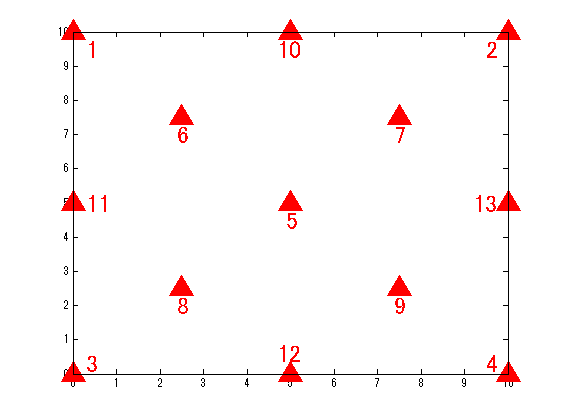}
&   &
\includegraphics[width= 3.5cm]{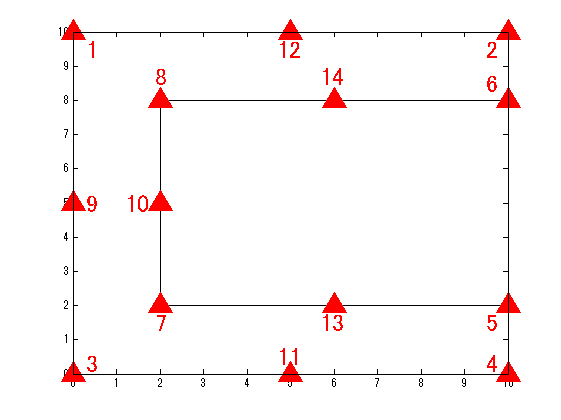}
\\
\footnotesize(a) Isotropic network &   & \footnotesize(b) Anisotropic network
\\	
\end{tabular}
\caption{Illustration of fixed anchor node locations.}
\label{fig:fixed}
\end{figure}


\section{Performance evaluation and analysis
\label{performance}}

In this section, we demonstrate using computer simulations the effectiveness of kHopLoc, and also compare to the original DV-hop\cite{niculescu2003dv} algorithm and a recent improved variant employing an approximate shortest path between nodes \cite{lee2014multihop} which we will refer to as the ASP algorithm for short.
The ASP algorithm is also a multihop range-free localization algorithm that tolerates network anisotropy with a small number of anchors. 
A detoured path detection is proposed which measures the deviation in the hop count between the direct and shortest paths of a node pair. 
A novel distance estimation method is introduced to approximate the shortest path based on the path deviation and to estimate their Euclidean distance by taking into account the extent of the detour of the approximate shortest path.

To evaluate the performance of the kHopLoc in isotropic networks and anisotropic networks, we deploy sensor nodes randomly in a $10 \times 10$ square-shaped region (such that $A=100$) and in a $10 \times 10$ C-shaped of width $2$ (such that $A=52$), where communication probabilities in isotropic networks and anisotropic networks are assumed to follow Raleigh fading communication model and QUDG communication model, respectively. 
In Rayleigh fading communication model, we assume the path loss exponent to be $\eta=2$ and parameter $\beta=1$ in formula \eqref{connectedness}. 
In the QUDG communication model, we assume the maximum connection distance $d_{max}= 1$, and the degree of irregularity $DOI= 1.5$. 
Moreover, to study the characteristic localization errors we simulate results in both isotropic and anisotropic WSN deployment regions with variable  $a)$ anchor node positions (fixed and random), $b)$ number of anchor nodes, and $c)$ node densities.
Communication and computation overhead costs are also compared and discussed.

\subsection{Localization error}

\begin{figure}[t]
\centering
\includegraphics[width=8.9cm]{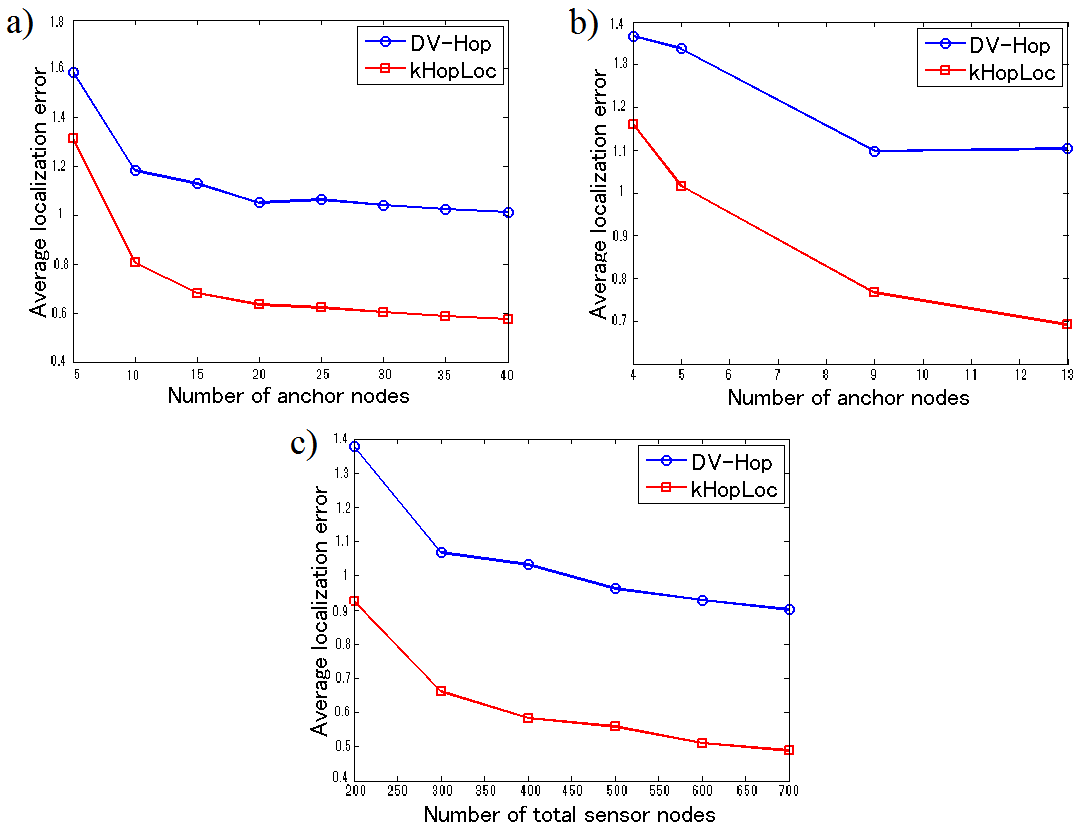}
\caption{Average localization error of DV-hop and kHopLoc in isotropic networks with random target node locations. 
a) $N=300$ with random anchor node locations.
b) $N=300$ with fixed anchor node locations as in Fig. \ref{fig:fixed}(a). 
c) $M=13$ fixed anchor node locations and $N\in[200,700]$.}
\label{fig:error-square}
\end{figure}

Fig. \ref{fig:error-square} depicts the average localization errors of DV-hop algorithm and kHopLoc by varying the number of anchor nodes and total sensor nodes in a square-shaped region under Rayleigh fading communication model.
In all cases, the error based on kHopLoc is significantly smaller than that of the  DV-hop algorithm with performance gains ranging between $20$ and $40\%$. 
Generally, the average localization error decreases with number of anchor nodes. 
Fixed anchor nodes (at strategic locations as in Fig. \ref{fig:fixed}(a)) in general provides better performance than randomizing anchor nodes, especially when the number of anchor nodes is small.
This is clearly due to the controlled avoidance of overlaps between anchor nodes which can lead to duplicate mutual information. 

\begin{figure}[t]
\centering
\includegraphics[width= 8.5cm]{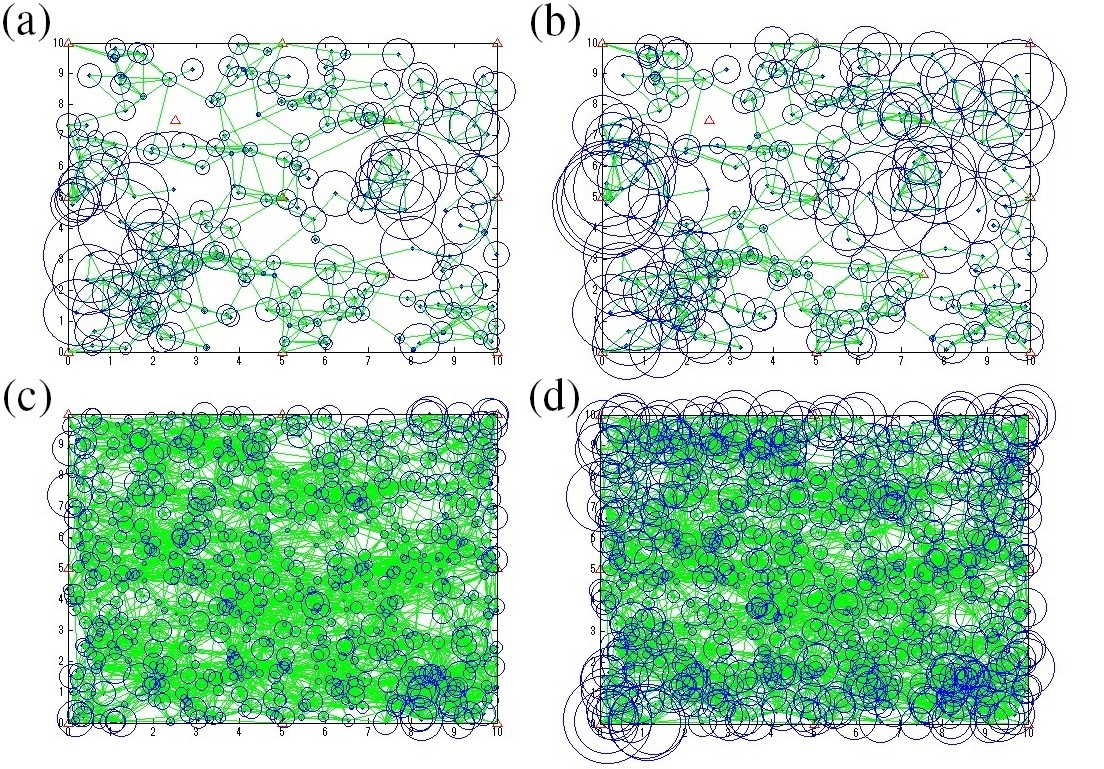}
\caption{Example isotropic WSN topologies and localization errors shown as disks. 
In all cases there are $M=13$ fixed anchor nodes as in Fig \ref{fig:fixed}(a).
Panels (a) and (c) use kHopLoc and $N=200 ,700$ sensor nodes.
Panels (b) and (d) use DV-hop algorithm and $N=200 ,700$ sensor nodes.
It is clear that kHopLoc has smaller localization errors, particularly near the boundary.}
\label{fig:error-by-case}
\end{figure}

Fig. \ref{fig:error-by-case} illustrates two example network topologies (a sparse regime at $N=200$, and a dense regime at $N=700$), and highlights the localization error of each node.
There are exactly $M=13$ fixed anchor nodes in all cases. 
The green lines connecting the nodes and blue circles describe communication links and localization errors in which the radius are proportional to the localization errors. 
In kHopLoc (panels (a) and (c)), errors of nodes having few links tend to be big, thus the average error decreases in the dense regime of $N=700$. 
On the other hand, in the DV-hop algorithm, localization errors of nodes near the border tend to be significantly larger.

\begin{figure}[t]
\centering
\includegraphics[width=8.9cm]{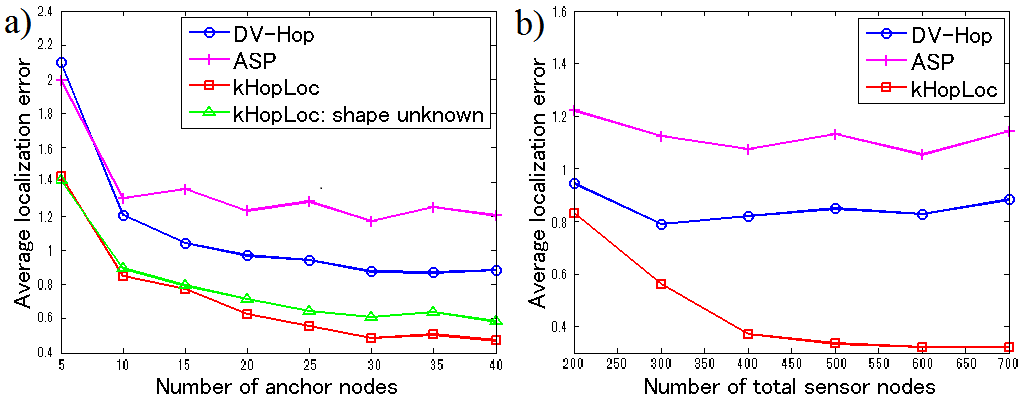}
\caption{Average localization error of DV-hop, ASP, and kHopLoc in anisotropic networks with random target node locations. 
a) $N=300$ with random anchor node locations.
b) $M=14$ fixed anchor node locations (as in Fig. \ref{fig:fixed}(b)) and $N\in[200,700]$.}
\label{fig:error-Cshape}
\end{figure}

Fig. \ref{fig:error-Cshape} depicts the average localization error of the DV-hop algorithm \cite{chen2012improved}, the ASP algorithm \cite{lee2014multihop}, and kHopLoc, for varying number of anchor nodes and total sensor nodes in C-shaped anisotropic region under the QUDG communication model. 
In all cases, the error based on kHopLoc is smaller than the other algorithms.
Notice that there are two result curves for kHopLoc: when the density and shape of the deployment region are known (green curve), and when both the density and shape of the region are unknown (red curve). 
The performance of the case when the region shape is known is better than the other case since the Monte Carlo simulation phase of the algorithm over a known shape region produces more precise distributions and thus results for the multihop connection probability than when the region is assumed to be a square. 
Significantly, it is worth noting that the average localization error due to kHopLoc continues to decrease for the other two algorithms seems to saturate after $300$ nodes. 
This demonstrates the benefits of using the full statistical hop-distribution generated during the first step of our algorithm (see section \ref{k-hop-probability}) rather than just first order statistics such as the mean one-hop Euclidean distance.

\begin{figure}[t]
\centering
\includegraphics[scale=0.32]{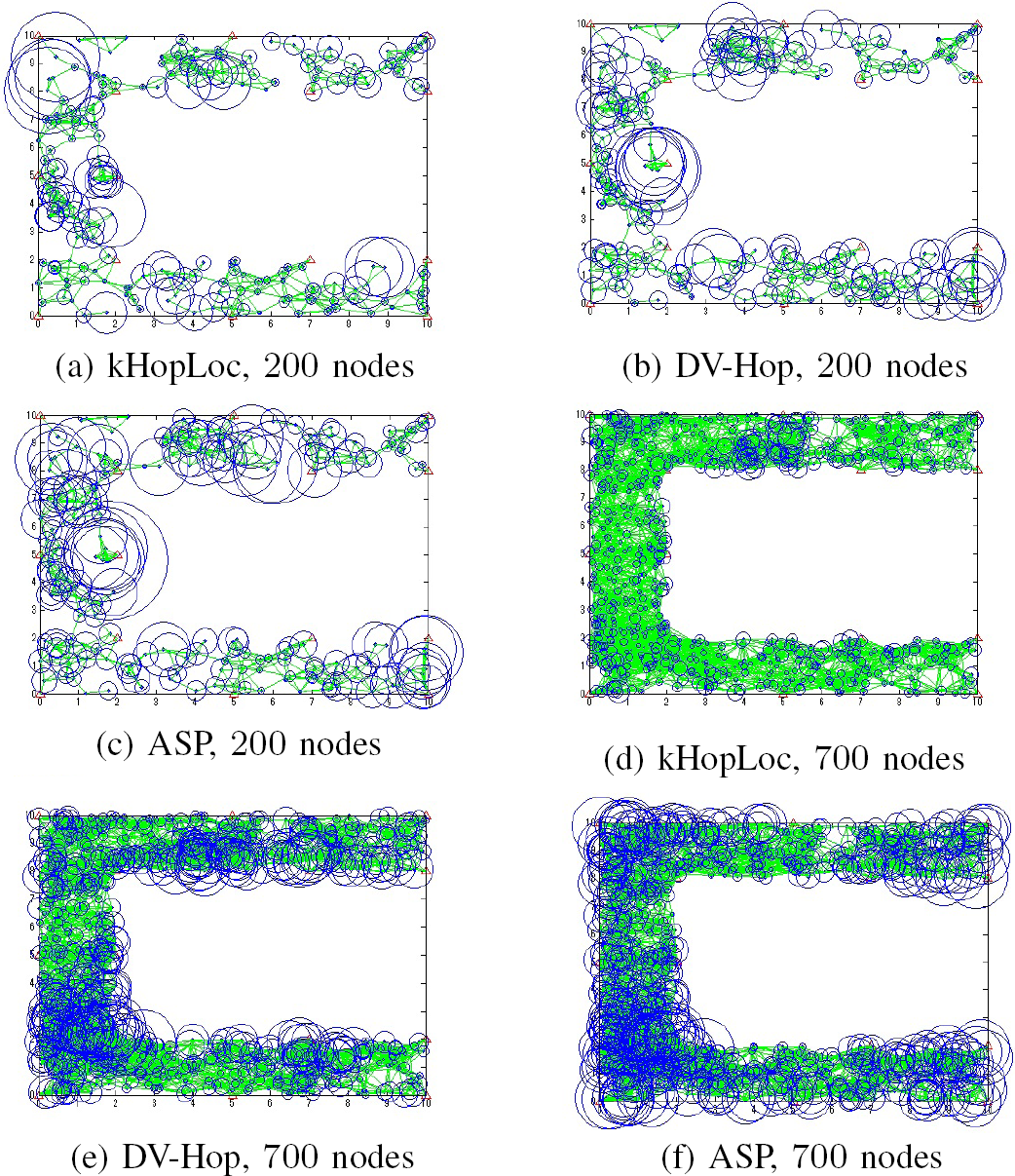}
\caption{Example anisotropic WSN topologies and localization errors shown as disks. 
In all cases there are $M=14$ fixed anchor nodes as in Fig \ref{fig:fixed}(b).}
\label{fig:error-by-case-anisotropic}
\end{figure}

Fig. \ref{fig:error-by-case-anisotropic} illustrates network topologies (a sparse regime at $N=200$, and a dense regime at $N=700$), and highlights the localization errors of each node. There are exactly $14$ fixed anchor nodes in the anisotropic C-shaped WSN deployment regions as shown in Fig. \ref{fig:fixed}(b). 
The green lines connecting the nodes and blue circles describe communication links and localization errors in which the radius are proportional to the localization errors. 
Similar to isotropic network, in kHopLoc, the main source of localization errors is due to nodes having few one-hop links.  
The reason for this is that nodes with fewer links tend to require a larger than average hop-count to reach anchors, thus making it difficult to estimate these node locations accurately.
As can be seen, this improves significantly in the dense regime.
On the other hand, localization errors in the other two algorithms seem not to improve with node density. 
The reason for this is that these  algorithms suffer from inaccurate inter-node distance estimations. 

\subsection{Overhead Analysis \label{overheads}}

This section discusses the communication and computational costs of the three localizations algorithms under investigation: DV-hop, ASP, and kHopLoc.

\subsubsection{Computational cost}
For calculating location of one target node, DV-hop algorithm costs $\mathcal{O}(M^8)$ if using normal matrix multiplication.
This cost comes from the matrix operations necessary to estimated the least square error \cite{wu2011improved}. 
The ASP algorithm costs consist of additional computation costs of $\mathcal{O}(\Delta)$ (where $\Delta$ is the number of tagged partitions for calculating Riemann sum) for calculating node density and $M(M-1)/2\cdot \mathcal{O}(\epsilon^{-3/2})$ for estimating distances between the target node and $M$ anchor nodes, where $M(M-1)/2$ is number of compound shortest paths, and $\mathcal{O}(\epsilon^{-3/2})$ (in which $\epsilon$ is the upper bound of the norm of the gradient) is the number of iterations of Newton method \cite{cartis2010complexity} for calculating the optimal central angles of the virtual holes \cite{lee2014multihop}. 
Consequently, the total computation complexity of the ASP algorithm is $\mathcal{O}(M^8+ \Delta+ M^2\cdot \epsilon^{-3/2})$.

On the other hand, the cost of kHopLoc consists of Monte Carlo simulation cost and the MLE localization cost. 
The former computational cost requires generating the random network, finding the shortest $k$-hop paths and fitting. 
Thus, the cost comes up to $\mathcal{O}(I\cdot N^3+ K\cdot L \cdot \epsilon^{-3/2})$, where $I$ is number of Monte Carlo iterations and $\mathcal{O}(N^3)$ is the cost for calculating all-pairs shortest paths if using for example the Floyd Warshall algorithm \cite{korte2002combinatorial}, and  $\mathcal{O}(K\cdot L \cdot \epsilon^{-3/2})$ is the fitting cost (see \eqref{fitting}) if using Newton method.
However, this computation can be done just once at a central node (or some server e.g. at the gateway) and then flooded through the network thus incurring an additional communication cost of $\mathcal{O}(N)$. Or otherwise, it is done before deploying the sensor nodes and then derived parameters ($A(k), B(k)$) are included into each sensor nodes, thus no additional communication cost occurs.

The MLE cost is $\mathcal{O}(M\cdot \epsilon^{-3/2})$, where $\mathcal{O}(\epsilon^{-3/2})$ is the number of iteration of Newton method and $\mathcal{O}(M)$ is cost for calculating the value of function \eqref{maxlikelihood4} in each iteration. 
Therefore the total cost amounts to $\mathcal{O}(I\cdot N^3+M\cdot \epsilon^{-3/2})$
Obviously, the MLE localization cost is smaller than that of the ASP algorithm.

\subsubsection{Communication cost}
The communication costs of the DV-hop algorithm and the ASP algorithm are bounded by $2\mathcal{O}\pr{M(N-M)}$, where $M$ and $N-M$ are number of anchor and target nodes respectively. 
This is because these algorithms perform flooding twice - first for the minimum hop count estimation, and second to broadcast the average one-hop distance.

The communication cost of kHopLoc however is mainly due to the initial hop count calculation  giving $\mathcal{O}(M(N-M))$ (i.e. similar to \cite{wang2009range}). 
When the density $\rho$ is unknown, additional communication costs of $2\mathcal{O}(N)$ can be incurred, in which  $\mathcal{O}(N)$ occurs when the nodes pass theirs $1$-hop neighbour number to the central node. 
After the central processor runs the Monte Carlo simulations and fits the said distributions, it then broadcast the results (parameters $A(k), B(k)$) to all nodes, thus costing another $\mathcal{O}(N)$.
The latter may lead to bandwidth issues in terms of the amount and resolution of the feedback information being flooded, thus suggesting a trade-off between communication overheads and fitting and localization accuracy.


\section{Conclusions \label{conc}}

We have proposed a maximum likelihood based multihop localization algorithm called kHopLoc for use in wireless sensor networks (WSNs). 
The main advantage of the algorithm is the use of a Monte Carlo initial training phase to generate the multihop connection probability density functions.
These are then used to build likelihood functions whose maxima estimate each target node location. 
Since the algorithm uses full statistical information for the multihop connection probabilities, localization results are significantly (about $20-40\%$) more accurate for both in isotropic and anisotropic networks. 
We have validated these results through computer simulations and discussed how and why some localization errors appear.
Finally, we have discussed the communication and computational costs of kHopLoc compared to conventional ones.
Moreover, like most range-free algorithms, kHopLoc can be used in conjunction with GPS and/or range-based localization schemes to improve performance and energy consumption in WSNs.
In the future, we aim to address the outstanding issue of localization in WSNs with non-uniform node deployments.

\section*{Acknowledgments}

This work has been supported by Corporate Research \& Development Center, Toshiba Corporation and Toshiba Telecommunications Research Laboratory.



\bibliographystyle{ieeetr}
\bibliography{mybib}

\end{document}